\documentclass[twocolumn]{openjournal}
\usepackage{graphicx, amsmath}
\usepackage[colorlinks=true,urlcolor=blue,
anchorcolor=blue,citecolor=blue,filecolor=blue,
linkcolor=blue,menucolor=blue,
linktocpage=true,pdfa=true]{hyperref}
\newcommand{\fig}[1]{\includegraphics[width=.47\textwidth]{figs/#1}}
\newcommand{\fign}[2]{\includegraphics[width=#1\textwidth]{figs/#2}}
\newcommand{\offline}{\mbox{$\overline{\rm Off}$\hspace{.05em}\raisebox{.3ex}{$\underline{\rm line}$}}}
\usepackage{orcidlink}

\begin{document}

\title{First Detection of Extensive Air Showers Using a Small-Aperture
Fluorescence Telescope
}

\author{M.~Zotov\orcidlink{0000-0003-0334-2367}}
\email{zotov@eas.sinp.msu.ru}
\affiliation{D.V.~Skobelstyn Institue of Nuclear Physics, M.V.~Lomonosov Moscow State University, Moscow 119991, Russia}

\author{A.~Trusov}
\affiliation{Faculty of Physics of M.V.~Lomonosov Moscow State University, Moscow 119991, Russia}

\author{P.~Klimov\orcidlink{0000-0001-9815-6123}}
\affiliation{D.V.~Skobelstyn Institue of Nuclear Physics, M.V.~Lomonosov Moscow State University, Moscow 119991, Russia}
\affiliation{Faculty of Physics of M.V.~Lomonosov Moscow State University, Moscow 119991, Russia}

\author{K.~Asatryan}
\affiliation{Cosmic Ray Department, Yerevan Physics Institute, Yerevan 0036, Armenia}

\author{A.~Belov\orcidlink{0000-0002-8565-6409}}
\affiliation{Faculty of Physics of M.V.~Lomonosov Moscow State University, Moscow 119991, Russia}
\affiliation{D.V.~Skobelstyn Institue of Nuclear Physics, M.V.~Lomonosov Moscow State University, Moscow 119991, Russia}

\author{G.~Gabaryan}
\affiliation{Cosmic Ray Department, Yerevan Physics Institute, Yerevan 0036, Armenia}

\author{V.~Kudryavtsev\orcidlink{0009-0003-2685-5330}}
\affiliation{Faculty of Physics of M.V.~Lomonosov Moscow State University, Moscow 119991, Russia}
\affiliation{D.V.~Skobelstyn Institue of Nuclear Physics, M.V.~Lomonosov Moscow State University, Moscow 119991, Russia}

\author{A.~Murashov}
\affiliation{D.V.~Skobelstyn Institue of Nuclear Physics, M.V.~Lomonosov Moscow State University, Moscow 119991, Russia}

\begin{abstract}

	We report on the detection of extensive air showers (EAS)
	generated by  ultra-high-energy cosmic rays using a small-aperture
	fluorescence telescope (FT) deployed at the Mount Aragats
	high-altitude research station.  The instrument is equipped with two
	Fresnel lenses with a diameter of 25~cm and operates with a
	2.625~$\mu$s time resolution. To our knowledge, this represents the
	first-ever observation of EAS achieved with an FT of such a compact
	aperture. To isolate shower events from the observational data, we
	implemented two independent event selection pipelines: a conventional
	cut-based analysis and a deep learning approach utilizing neural
	networks. Both algorithms successfully identified over 15
	high-confidence EAS tracks from data acquired during clear, moonless
	nights. We present selected event topologies and detail the
	background rejection methodology employed to discriminate true shower
	tracks from spurious signals mimicking EAS signatures.
	These results provide an important proof-of-concept for the
	advancement of fluorescence detection techniques, demonstrating their
	viability for forthcoming ground-based and space-borne missions.
	Future efforts will focus on primary energy reconstruction utilizing
	a previously developed neural-network framework.

\end{abstract}


\maketitle

\section{Introduction}

Upon entering the Earth's atmosphere, cosmic rays (CRs) induce secondary
particle cascades known as extensive air showers (EAS). During the
development of an EAS, the interaction of secondary charged particles
with atmospheric nitrogen molecules yields fluorescence emission. This
emission can be observed in the near-ultraviolet (UV) band during clear,
moonless nights using dedicated telescope systems~\citep{Greisen-1965,
flyseye-1975}. Because the number of emitted photons is directly
proportional to the energy deposited by the shower, observing this
emission allows for the reconstruction of the primary particle's arrival
direction, as well as the estimation of its energy and mass composition.
Crucially, energy estimations derived from fluorescence measurements are
largely independent of hadronic interaction models, unlike those
obtained from various surface detector arrays such as water
Cherenkov detectors, scintillation counters, and others. Consequently,
fluorescence telescopes (FTs) have become a pivotal component of modern
cosmic-ray observatories operating at energies above approximately
10~PeV ($10^{16}$~eV), and particularly above 1~EeV ($10^{18}$~eV)~---
the ultra-high-energy (UHE) regime~\citep{auger-fd, ta-fd, lhaaso, heat}.

Conventional fluorescence telescopes are highly complex and bespoke
instruments, employing large-area mirrors (approximately 10~m$^2$ in the
Auger~\citep{auger-fd} and Telescope Array (TA)~\citep{ta-fd}
experiments) and arrays of expensive photomultiplier tubes (PMTs). These
constraints have motivated the development of more compact and
cost-effective FTs with smaller apertures. For instance, in 2014, the
international JEM-EUSO collaboration developed EUSO-TA, a refracting
telescope featuring an optical system composed of two 1~m diameter
Fresnel lenses~\citep{eusota-2015, eusota-2018}. This instrument was
designed as a prototype for the large-scale JEM-EUSO space
mission~\citep{jemeuso-mission, jemeuso-instrument} to validate its
optical design and electronics. The telescope was deployed at the
Telescope Array site in Utah, USA (hence the ``TA'' designation).
EUSO-TA lacked an independent trigger system; instead, it operated using
trigger information provided by a TA fluorescence detector {located
at Black Rock Mesa site and pointed in the same direction}. Although
EUSO-TA was not primarily intended for dedicated physical observations,
it successfully recorded 9 EAS events induced by CRs with energies
ranging from approximately 0.3 to 3~EeV during several observation runs
in 2015~\citep{eusota-2024}.

Concurrently, two Japanese projects proposing FTs with apertures of
approximately 2~m$^2$ emerged: FAST~\citep{fast} and
CRAFFT~\citep{crafft-2025}. Unlike EUSO-TA, both initiatives were
specifically conceived to demonstrate the feasibility of employing
small-aperture FTs for UHECR observations in ground-based experiments.
Currently, several prototypes of these telescopes are undergoing field
testing at the Auger and TA observatory sites. These modules are being
considered for potential integration into the planned GCOS
experiment~\citep{gcos-2025}.

In 2023, the SINP MSU developed a fluorescence telescope equipped with
25~cm diameter lenses based on the engineering model of the
Russian-Italian ``UV atmosphere'' (Mini-EUSO) space
telescope~\citep{mini-euso}. This instrument was originally designed for
application in the PAIPS experiment, which aims to study auroral
emissions in the UV band~\citep{paips-2025}. During the summer months of
2024 and 2025, as part of a collaborative project with Yerevan State
University (Republic of Armenia), two observation campaigns were
conducted at Mount Aragats with the main goal of studying UV emission
from energetic thunderstorm processes. During these campaigns, EAS
tracks were detected for the first time using an FT with
such a compact aperture. This achievement opens up new prospects for the
development of small-aperture FTs. Such instruments could serve both as
prototypes for future space-borne missions (e.g., within the ERA project
framework~\citep{ERA-iscra2025}) and as sub-detectors for the
forthcoming TAIGA-100 ground-based observatory~\citep{taiga-100},
providing independent calibration and complementing the surface detector
array.

In this paper, we describe the Small-Aperture Fluorescence Telescope
(SAFT) utilized for these observations and outline the methodologies
employed to search for EAS tracks within its data set. We
present a selection of the detected events and discuss potential
directions for future research.

\section{Small-Aperture Fluorescence Telescope}

As previously mentioned, the small-aperture fluorescence telescope
employed in this study is based on the engineering (pre-production)
model of the Mini-EUSO instrument (``UV atmosphere'' in Russian Space
Program). Similar to Mini-EUSO, the optical system consists of two
double-sided Fresnel lenses with a diameter of 25~cm and a flat focal
surface.  {The overall length of the optical system equals 300~mm.
The effective focal length is 150~mm, which results in the focal number
of~0.6. The lenses were manufactured from UV-transparent polymethyl
methacrylate (PMMA-000, MITSUBISHI RAYON).  The thickness of each lens
is 11~mm.  The optical system of two Fresnel lenses was chosen to
achieve a wide field of view with a relatively light and compact design
suitable for space application.  A detailed description of the optical
system of Mini-EUSO can be found in~\citet{minieuso-optics}.}

\begin{figure*}[!ht]
	\fig{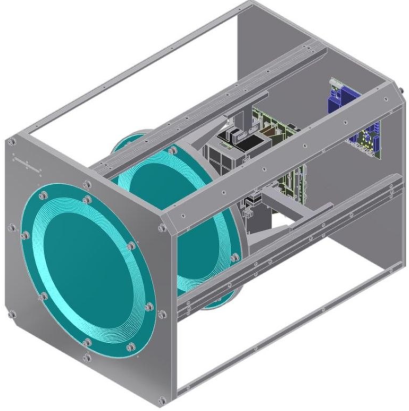}\quad\fign{.51}{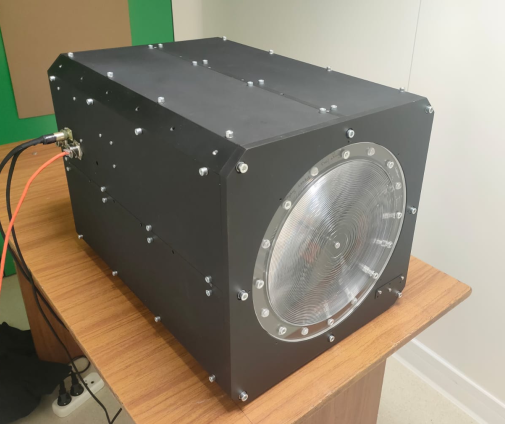}

	 \caption{The small-aperture fluorescence telescope (SAFT) deployed
	 at Mount Aragats.  Top: 3D schematic view of the instrument. Bottom:
	 photograph of the fully assembled detector.
	 }

	 \medskip
    \label{fig:paips}
\end{figure*}

Unlike Mini-EUSO, the focal surface of the SAFT is built of 12 Hamamatsu
multi-anode photomultiplier tubes (MAPMTs, {type R11265-M64}).  This
configuration results in a matrix of $48\times16$ pixels, compared to
the $48\times48$ layout of the original Mini-EUSO.  {Each pixel has
the size of $2.9\times2.9$~mm$^2$.} The MAPMTs are highly sensitive and
{are configured to operate} in the single-photon counting mode.
{Each MAPMT is covered with a 2-mm BG3 UV filter with
anti-reflective coating.}  Accounting for the dead spaces between
pixels, the instrument yields a field of view (FoV) of approximately
$44^\circ\times15^\circ$.  {Tests of the optical system revealed
that photon collection efficiency in a circle of 3.3~mm diameter (i.e.,
of equivalent circular area) is about 45\% in the whole FoV up to field
angles $\sim20^\circ$.} Figure~\ref{fig:paips} presents a 3D schematic
rendering of the SAFT alongside a photograph of the fully assembled
instrument.

{The SAFT was calibrated at Lomonosov Moscow State University before
the observation campaigns. However, a protecting glass window was needed
in front of the instrument at the Aragats station because it operated in
different weather conditions.  The window affected the sensitivity of
the detector but dedicated tests could not be performed due to the lack
of necessary equipment at the site.  Because of this, we only show
photon counts instead of the absolute photon flux in what follows.}

The data acquisition system operates simultaneously with two distinct
time resolutions. In the ``D1'' mode, the time duration of a single
frame (time bin) equals 2.625~$\mu$s. Each triggered event record
encompasses 128 frames, resulting in a total readout window of
336~$\mu$s per event.  Event acquisition in this mode is governed by a
trigger logic similar to the one developed for the fluorescence
telescope of the EUSO-SPB2 stratospheric balloon
experiment~\citep{spb2-trigger}.  In the second mode (``D3''), data are
recorded continuously, but the frame duration is integrated over
$2.625~\mu\text{s}\times128\times128\approx43$~ms. Given that EAS
detection necessitates a high sampling rate, all shower events discussed
hereafter were captured exclusively in the D1 mode.  {As long as the
SAFT was aimed for observing auroral emissions in the D3 mode,
a dedicated trigger for registering EAS from the ground was not
implemented.}

It should be noted that {time resolution of the order of 1~ms
perfectly fits observations of extensive air showers from an orbit with
a height of 400--500~km~\citep{jemeuso-instrument} but is sub-optimal}
for ground-based EAS observations.  For context, the fluorescence
telescopes of the Auger and TA observatories operate with a 100~ns
sampling rate, which allows for a detailed recording of the longitudinal
shower development. The SAFT inherits the 2.625~$\mu$s resolution
directly from the front-end electronics originally designed for the
space-borne mission.  {As a result, one cannot expect registering
evolution of an EAS unless it has an arrival direction close to the line
of sight of the detector. Recall that EUSO-TA, which had the time
resolution of 2.5~$\mu$s, recorded events within at most two time
frames~\citep{eusota-2024}.}

\begin{figure*}[!ht]
    \centering
	 \fign{.7}{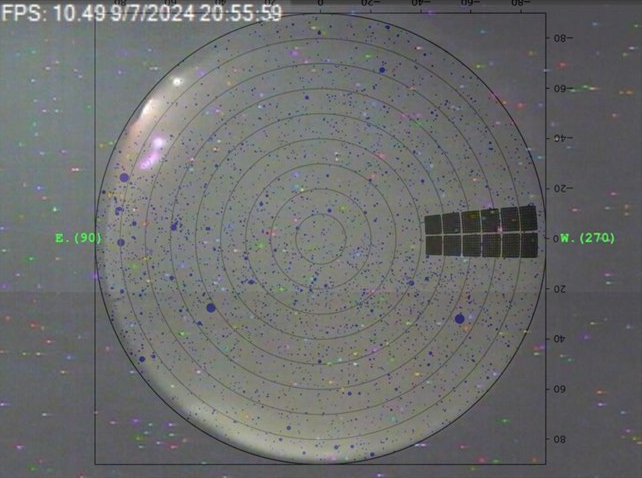}
    \caption{All-sky camera image taken at the Aragats Research Station.
	 North is at the top.
	 The black shaded area represents the {$48\times16$ pixel}
	 field of view of the SAFT.
    }
    \label{fig:paips-fov}
\end{figure*}

The instrument was deployed at the Aragats Research Station of the
Yerevan Physics Institute, located at an altitude of 3200~m above sea
level, corresponding to an atmospheric depth of approximately
700~g~cm$^{-2}$.  Observational campaigns were conducted exclusively
during the summer months, which allowed maximizing the number of clear,
moonless nights required for detecting faint fluorescence signals.  The
telescope was oriented at an elevation angle of $20^\circ$. The
dimensions and spatial orientation of the instrument's FoV are
illustrated on an all-sky camera
image\footnote{\url{http://crd.yerphi.am/Aragats\_Sky\_Monitoring}} in
Fig.~\ref{fig:paips-fov}.  There are 3 all-sky cameras at Aragats,
synchronized with other facilities, providing a continuous flow of
images with {a 1-minute time span}. Images from these cameras will be used
in subsequent sections to visualize the observational conditions during
the detection of specific tracks.

\section{EAS Track Search Methodology}

On the focal surface, the EAS fluorescence signal manifests as a
quasi-linear track. Its length, intensity, and spatial spread depend on
the primary particle energy and type, the event geometry relative to the
instrument, the point spread function (PSF) of the optical system, and
other factors. To identify such signals, we evaluated several
algorithms, initially validating them on simulated data. Because a
dedicated software simulator for the SAFT was not available at the time
of this study, we utilized a dataset generated using the
CONEX~\citep{conex} and EUSO-\offline~\citep{offline} frameworks,
originally developed for the EUSO-TA ground-based telescope.
{The initial data set consisted of more than 100 thousand proton showers
simulated in the energy range 1--6.6~EeV using the QGSJET-II-04 model of hadronic
interactions~\citep{qgsjet-2011}. Energies were distributed uniformly vs.\
$\log(E/\mathrm{eV})$ with the step of 0.02. Azimuth angles were
simulated uniformly while zenith angles varied from $0^\circ$ up
to $45^\circ$ with their number $\sim\cos\theta$. All shower cores were
positioned within the projection of the detector FoV on ground.}

{Since the focal surface of EUSO-TA consists of $48\times48$ pixels,
we needed to adapt the data} to match the focal surface
geometry of our instrument.
{Namely, we extracted events that contained tracks within
a $48\times16$ matrix, with at least 32 hit pixels and the amplitude
$\ge7$ (including the uniform background illumination).}
This simulated dataset was augmented with
real observational data consisting of track-free frames including
various examples of signals induced by electronic noise and instrumental
artifacts. Incorporating this real noise background significantly
reduced the false-positive rate during subsequent analysis.

The first evaluated approach is a conventional cluster-finding algorithm
based on merging signals from adjacent pixels evaluated against their
statistical baselines.

\begin{figure}[!ht]
	\centering
	\fig{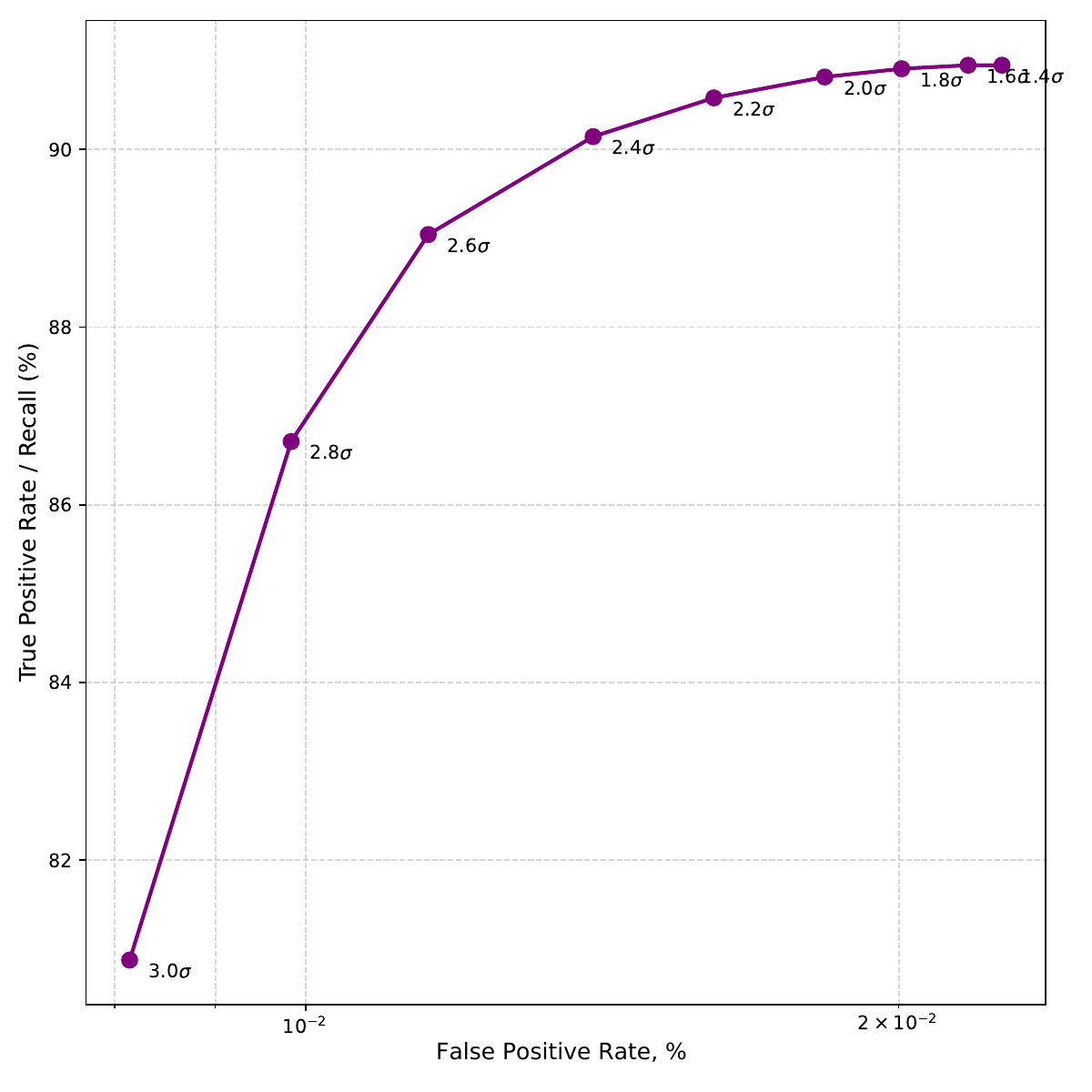}
	\caption{ROC curve illustrating the trade-off between the track
	detection efficiency (True Positive Rate/Recall) and the False
	Positive Rate as a function of the $\sigma$ threshold.}
	\label{fig:sigma}
\end{figure}

\begin{itemize}

	\item \textbf{Background estimation:} For each pixel, the running
		mean~(\(\mu\)) and standard deviation~(\(\sigma\)) of the signal
		were calculated over a sliding window of the 16 preceding time
		bins (frames).

   \item \textbf{Z-score normalization:} Each pixel's signal was
		normalized as follows:
		\[
			I_{\text{norm}} = \frac{I_\text{raw} - \mu}{\sigma},
		\]
		where \(I_\text{raw}\) is the raw pixel signal, \(I_\text{norm}\)
		is the normalized signal, \(\mu\) is the moving mean, and
		\(\sigma\) is the moving standard deviation.
		{This step can be considered as a kind of flat-fielding
		that partially compensates the lack of accurate calibration
		of the detector.}

		Pixels exhibiting a signal excess greater than 2\(\sigma\) above
		the mean were flagged as ``active''. For pixels with raw
		intensities below the mean (\(I_\text{raw} < \mu\)), the
		normalized signal was set to zero.

		The optimal threshold of $2\sigma$ was determined using the
		simulated dataset by balancing the signal detection efficiency
		(True Positive Rate or Recall) against the False Positive Rate
		(FPR). As demonstrated by the Receiver Operating Characteristic
		(ROC) curve analysis (Fig.~\ref{fig:sigma}), setting the threshold
		below $2\sigma$ results in a plateaued sensitivity, while the FPR
		continues to rise. Conversely, increasing the threshold beyond
		$2\sigma$ leads to a noticeable degradation in {signal
		detection} efficiency.

	\item {\sloppy \textbf{Cluster identification:} The algorithm exclusively
		processed frames containing connected active regions. Automated
		pixel clustering was performed using the \texttt{scipy.ndimage}
		library~\citep{scipy}, which identified connected components on the binary mask
		of active pixels (utilizing 8-connectivity within a 3$\times$3
		matrix). Subsequently, the pipeline rejected clusters containing a
		number of pixels below a predefined threshold.

		}

	\item \textbf{Spatial smoothing:} To further suppress false positives
		induced by instrumental artifacts or ambient light fluctuations, a
		spatial moving-average filter with a \(4\times4\) pixel kernel was
		applied to the normalized signal \(I_{\text{norm}}\). This
		smoothing compensates for the algorithm's inherent sensitivity to
		high-frequency background noise. Acting as a spatial low-pass
		filter, the moving average suppresses random single-pixel triggers
		and locally integrates the signal. Because a typical EAS track
		consists of dense signal clusters surrounded by a sparse optical
		halo, this smoothing effectively localizes these dense cores,
		facilitating robust track recognition.

	\item \textbf{Event {selection}:} An event was finally classified as a
		track candidate only if the maximum value of the smoothed signal
		exceeded a specific threshold. This condition ensures that the
		detected active pixels form a sufficiently dense topological
		group, rather than representing uncorrelated noise scattered
		across the focal surface.

\end{itemize}

\begin{figure*}
	\centering
	\fign{.65}{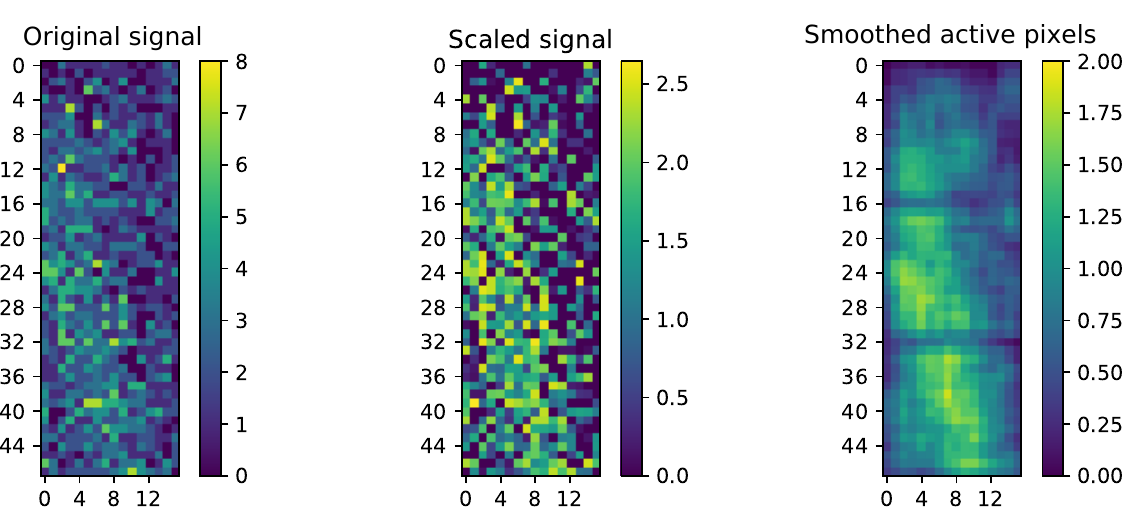}\\[5pt]
	\fign{.65}{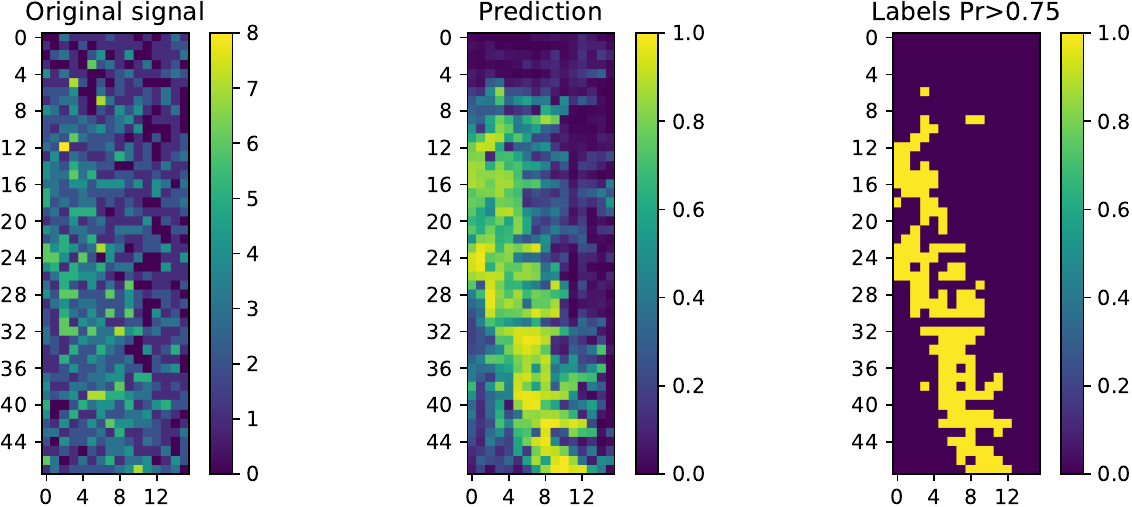}

	\caption{Example of simulated EAS track recognition using the Z-score
	normalization method and the convolutional encoder-decoder. Top row,
	from left to right: raw input signal {with colors indicating
	photon counts}, Z-score normalized signal, and
	spatially smoothed signal.  Bottom row, from left to right: the same
	raw input signal; pixel-wise probability map of belonging to a track
	as predicted by the CED; binary mask of active pixels selected with a
	probability threshold exceeding~0.75.
	{Here and below, numbers 0, 4,\dots,44 along y-axes of images
	and numbers 0, 4, 8, 12 along x-axes of images indicate numbers of
	rows and columns of pixels respectively. The top row (number~0)
	corresponds to the top of the FoV.}
	}

	\label{fig:track}
\end{figure*}

While this conventional method demonstrated high computational
efficiency and detection performance, it required further modifications
to mitigate a substantial number of false-positive {events} caused by
electromagnetic interference from external sources in the front-end
electronics.

The second approach developed for EAS track recognition relied on the
application of neural networks. The primary instrument acting as a
``software trigger'' for track detection in the SAFT data was a
convolutional neural network (CNN). The CNN performed a binary
classification task, sorting all incoming frames into two distinct
categories: those containing a track and all others (background or
noise).
Several lightweight architectures were evaluated, and it was
found that even a minimal configuration comprising a single
convolutional layer followed by a single fully connected layer achieved
satisfactory performance. Owing to its simplicity, this network
architecture imposed minimal computational overhead and demonstrated a
high processing speed, comparable to that of the conventional
cluster-finding algorithm.%
\footnote{{We have considered the possibility of implementing
a CNN-based trigger in the FPGA in order to eliminate the number of
false positives but this work is not finished yet. Its results will
be reported elsewhere.}}
{Training data sets consisted of 10 thousand events equally
split between those containing simulated tracks and ``noise''
extracted from the real data. Track events were extracted in a random fashion
from the initial simulated data set described above.
The CNN did not need preliminary flat-fielding to
find events with tracks, similar to the models suggested
by~\citet{minieuso-cnn-meteors} to recognize tracks of meteors in
Mini-EUSO data.}

Because we intend to address the task of primary energy reconstruction
for the detected EAS events in future work, we additionally implemented
a procedure to identify individual active pixels within the frames
pre-selected by the CNN trigger. To accomplish this, we employed a
convolutional encoder-decoder (CED) identical to the one
described in~\citet{reco1, Zotov-Trusov-2025}. Although this pixel-wise
segmentation step was not strictly necessary for the initial track
selection, it enabled the spatial visualization of the CNN's trigger
output, thereby significantly simplifying the final manual inspection of
the selected event sample. Figure~\ref{fig:track} presents an example of
a simulated EAS track processed by both the Z-score normalization method
and the CED approach.
A major advantage of the neural-network-based approach over the Z-score
normalization method is that it eliminates the need for manual tuning of
selection cuts to suppress false-positive triggers (which predominantly
arise from instrumental artifacts). By simply including representative
examples of these spurious events in the training dataset, the neural
networks implicitly learn to recognize and effectively reject them
during inference.
{We believe that performance of the neural-network-based approach
can be further improved as soon as we implement a model of the SAFT
in the EUSO-\offline{} framework.
This will allow us to prepare training data sets simulated specifically
for our instrument instead of the similar but much larger EUSO-TA
detector.}

\section{Track Events in the SAFT Data}

\begin{figure*}[!ht]
	\centering
	\fign{.56}{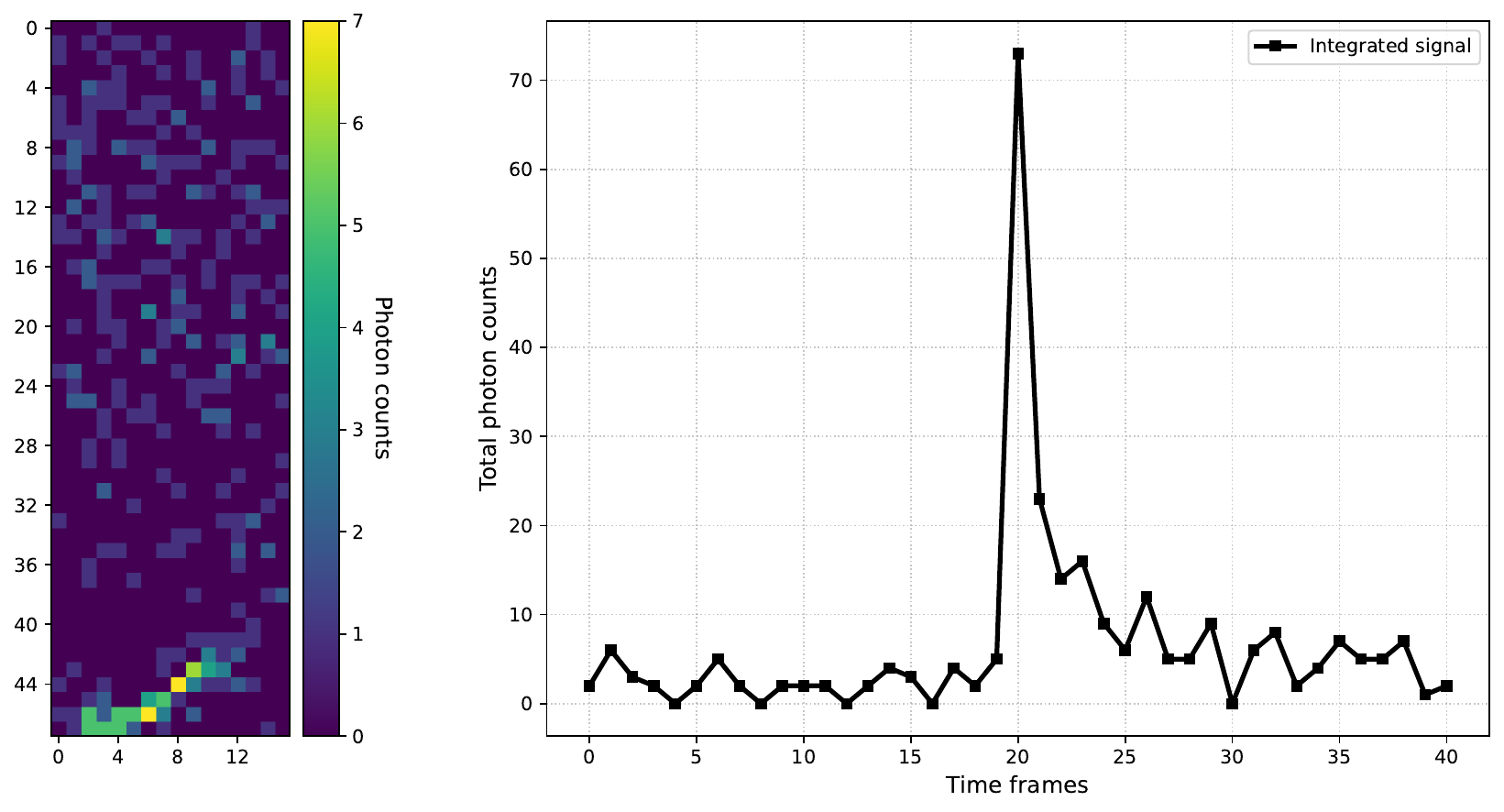}\\
	\fign{.56}{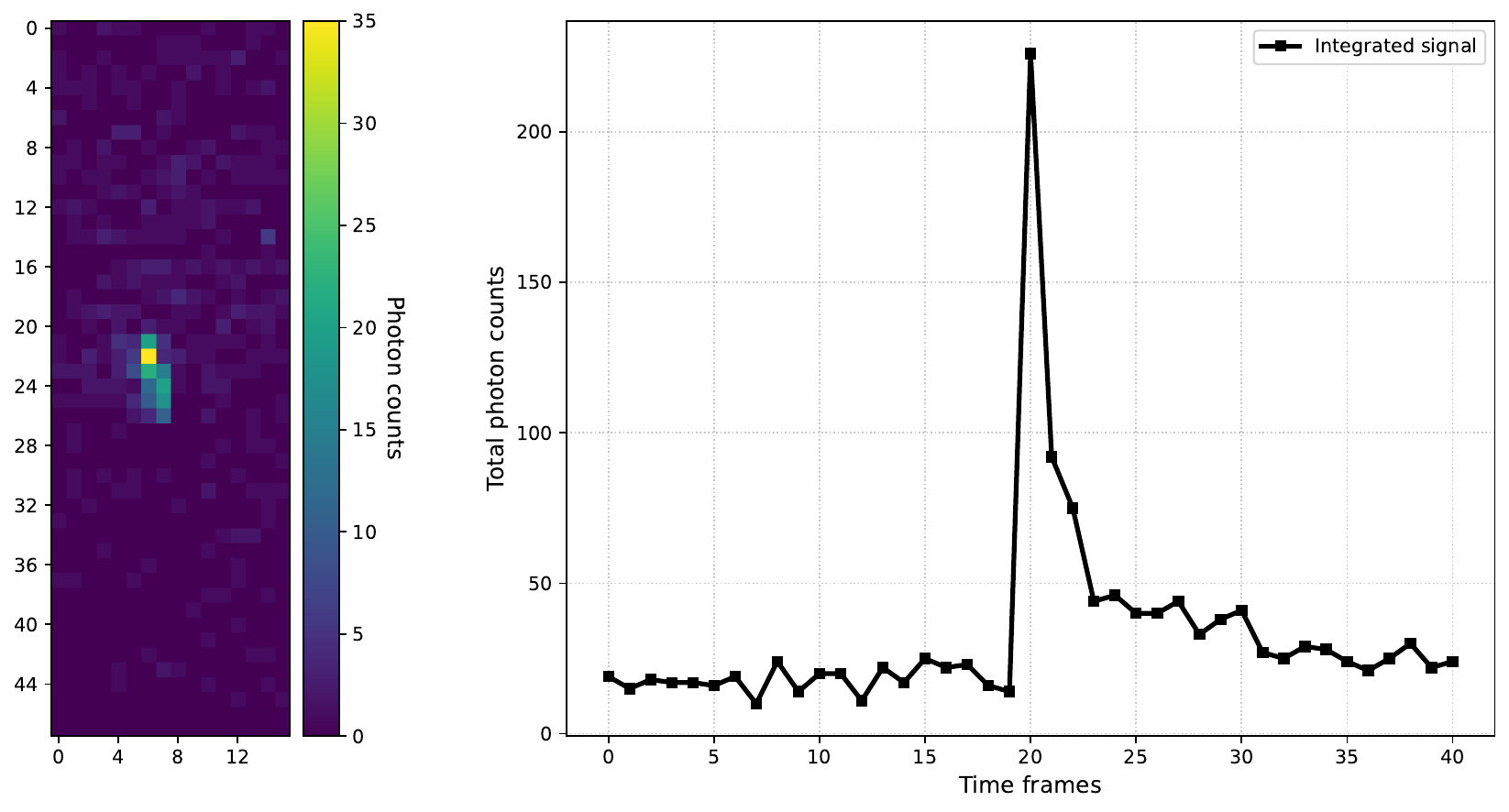}
	\caption{Examples of track-like events not induced by EAS fluorescence.
	The left panels display ``snapshots'' of the telescope's focal surface
	at the moment of maximum signal intensity.
	The right panels show integrated signals (light curves) over 40 time bins.
	}
	\label{fig:crhit}
\end{figure*}

{Approximately 17--20 thousand events were typically triggered in
the D1 mode during one hour of observations.}
The majority of track-like events identified in the SAFT data
exhibited morphologies similar to those shown in Fig.~\ref{fig:crhit}.
These events are characterized by a simultaneous signal spike within a
single time bin across a group of adjacent pixels arranged
quasi-linearly on the focal surface, followed by an exponential decay.
Analogous signal behavior was previously observed in the data of the
space-borne TUS telescope. According to simulation
studies~\citep{tus-tracks-izvran, tus-jcap-2017}, such tracks in the TUS
experiment were induced by direct hits of charged protons with energies
ranging from hundreds of MeV to several GeV on the photodetector. In
the present case, these tracks were most likely caused by secondary
particles comprising an EAS directly striking the instrument. The
hypothesis that the signal originated within the photodetector
or directly on one of the lenses, is supported by the complete absence
of optical blurring (PSF spreading) in the track images.

\begin{figure*}[h]
	\centering
	\fign{.56}{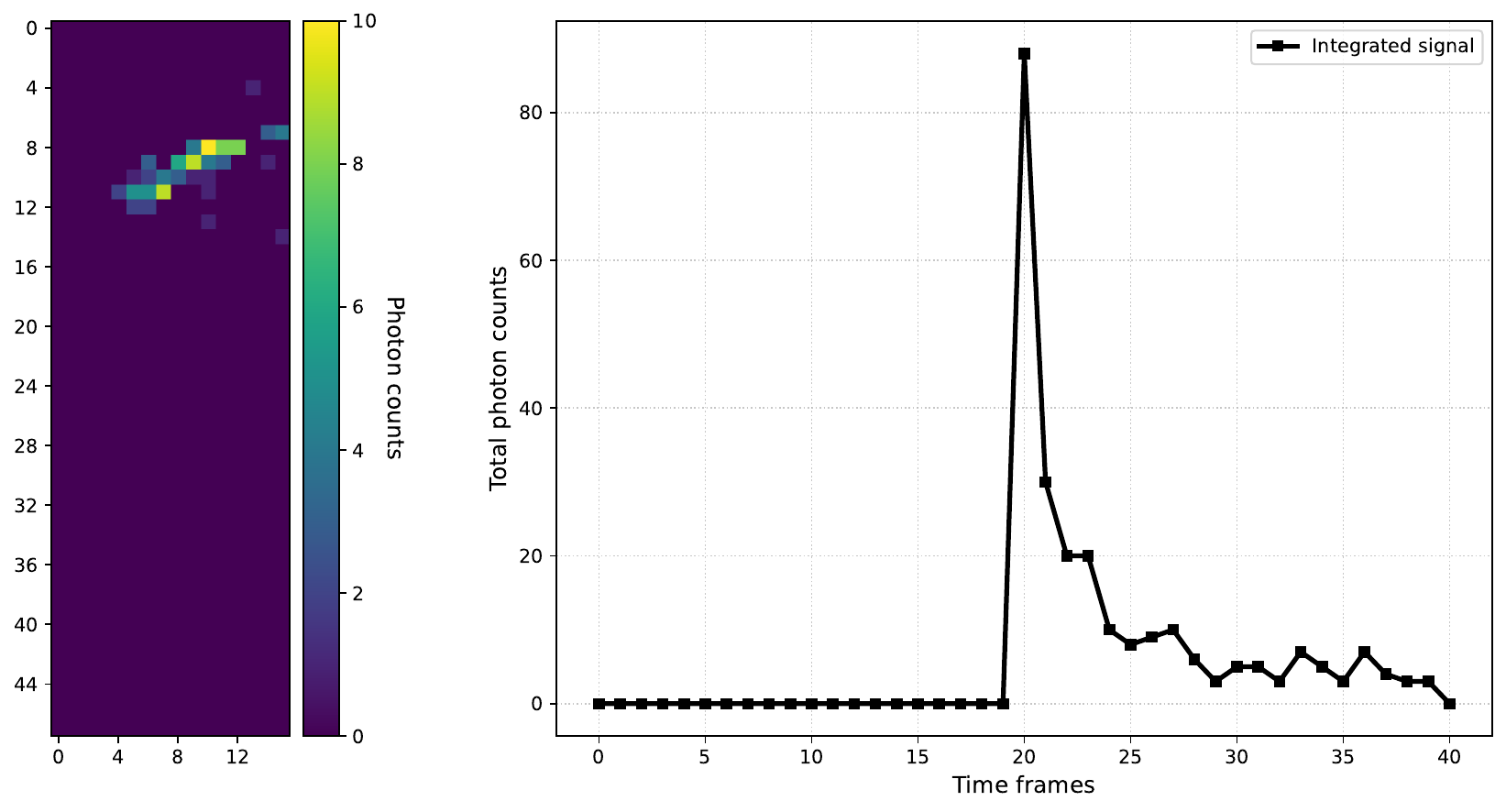}\\
	\fign{.56}{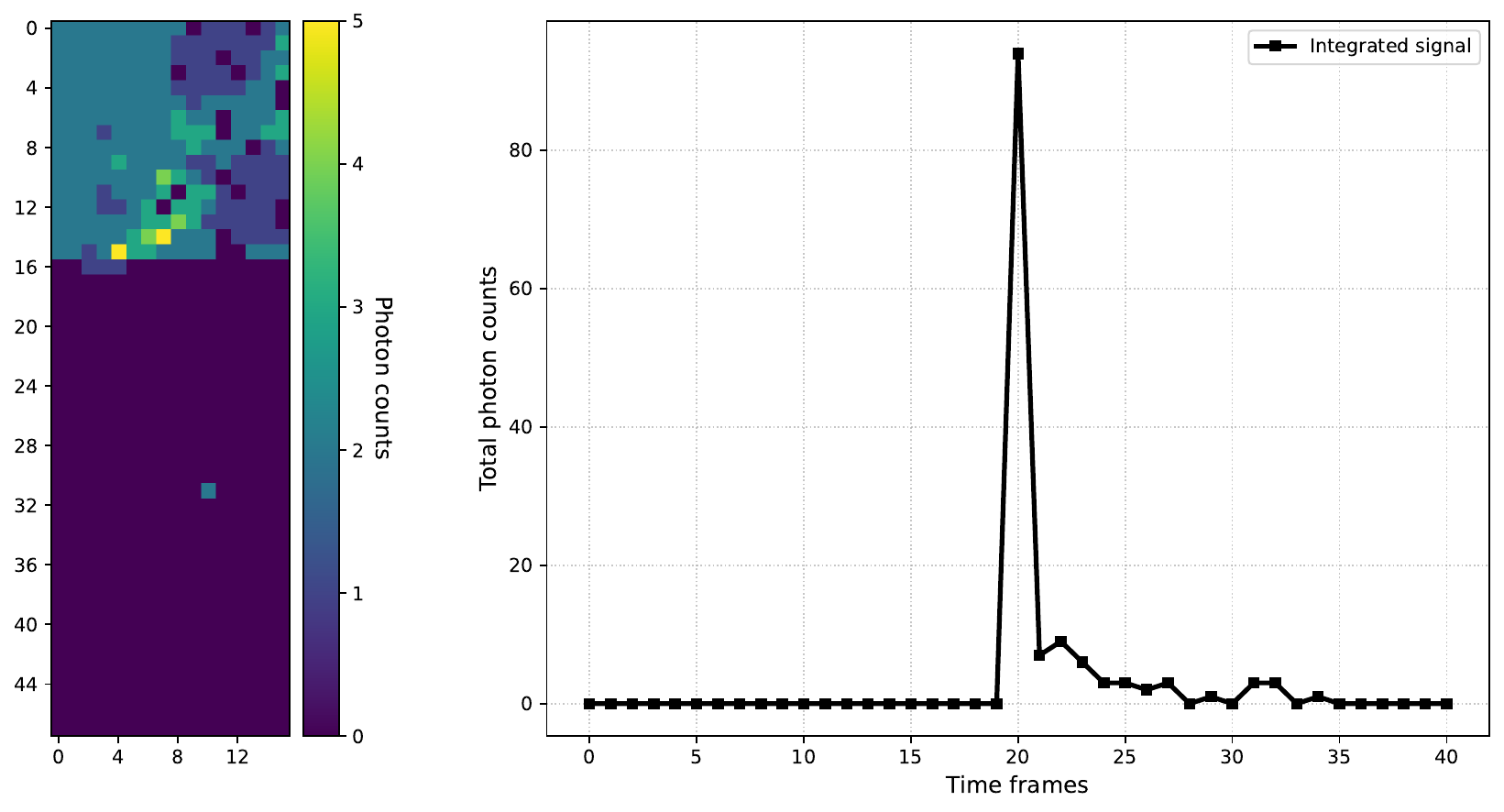}
	\caption{Examples of tracks recorded with the SAFT shutter
	closed. Left: snapshots of the focal surface at the signal maximum.
	Right: integrated signals (light curves) over 40 time bins.
	}
	\label{fig:blind}
\end{figure*}

{Our assumption that these signals do not originate from a distant
track-like source of illumination} was corroborated when identical
signal patterns were discovered in test data recorded with a closed
telescope shutter (dark frame data). Examples of these events are
presented in Fig.~\ref{fig:blind}. It is clearly evident that the
signals form quasi-linear tracks on the focal surface, while the
integrated light curve exhibits a sharp initial spike followed by an
exponential decay. Notably, in the second event, the decay rate is
significantly faster than in the first one.  Furthermore, in the second
case, an elevated signal appears across the entire elementary cell
defined as a $16\times16$ pixel block sharing a common high-voltage
supply, possibly due to the much higher luminosity of the signal.  An
exponential fit of the form $A\exp(-t/\tau)$ applied to the ``tails'' of
the integrated signals for these two event types revealed a decay time
constant of $\tau\approx4.2\dots5.2~\mu$s for the first type of tracks,
and $\tau\approx1.1~\mu$s for the second. The physical mechanism
responsible for the emergence of these two distinct track sub-types is
not yet fully understood. {It was suggested by one of the anonymous
reviewers of the first version of the paper that at least some of these
tracks can represent Cherenkov light from charged particles (mostly
muons) passing through the lenses. Figuring out whether one of the two
types of light curves appears due to direct hits of the photodetector
and another one is due to Cherenkov light in the lenses demands detailed
simulations and goes beyond the scope of the paper.} However, regardless
of their specific origin, tracks generated directly within the
photodetector—bypassing the telescope's optical system—are of no
interest for EAS studies and are therefore classified as spurious
background signals.

Consequently, during the subsequent data analysis, whenever a
track candidate was identified in the SAFT data, we rejected all events
exhibiting an exponentially decaying ``tail'' in the light curve and/or
triggering the entire elementary cell traversed by the track. We
acknowledge that this rejection criterion may be overly conservative. We
plan to refine this background rejection strategy in future studies
based on a detailed Monte Carlo simulation of the detector response.

\begin{figure*}[t]
	\centering
	\fign{.84}{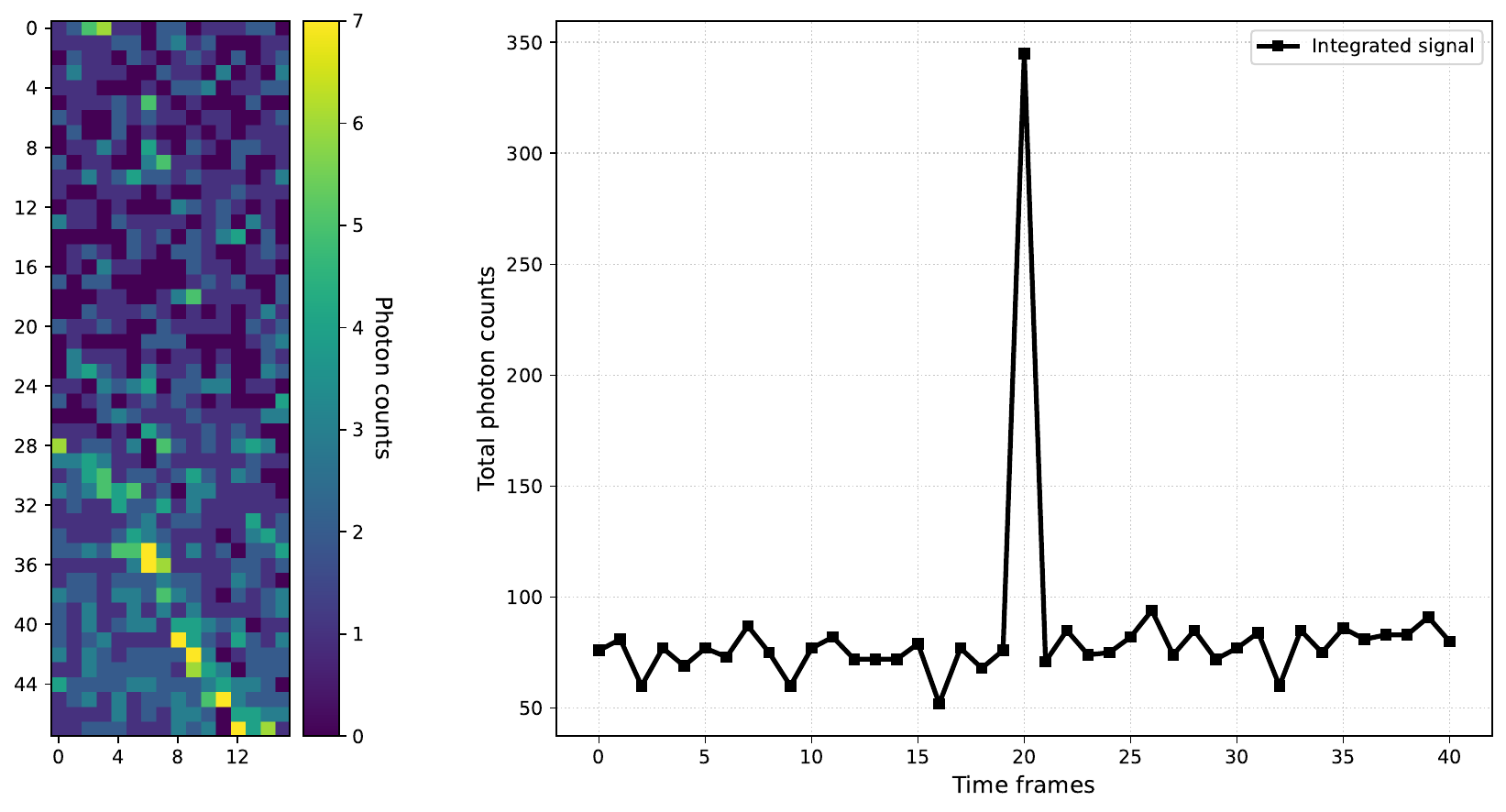}
	\fign{.82}{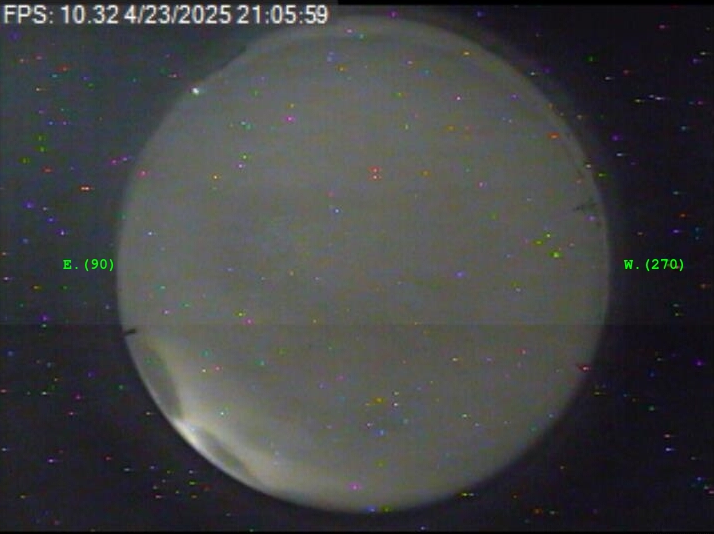}
	\caption{Top row: example of a genuine extensive air shower track recorded
	on 23.04.2025 at 21:06:18 UTC. The top panel displays the original SAFT signal
	{at the moment of registering the track, including the background} (left)
	and the integrated signal over the 40-bin readout window (right).
	Bottom: all-sky camera image taken on 23.04.2025 at 21:06~UTC.
	}
	\label{fig:250423}
\end{figure*}

The top row of Figure~\ref{fig:250423} presents an example of a track
that, in our assessment, was induced by actual EAS fluorescence
emission. It is clearly visible that the focal plane image of this track
is blurred (consistent with optical point spread), and its light curve
manifests as a single-bin spike devoid of any exponential afterglow.
This specific track was recorded on April 23, 2025, at 21:06:18 UTC,
four days prior to the new moon phase. The bottom panel of the same
figure displays an all-sky camera image captured a few seconds before
registration of the track.  The image confirms that the sky was
clear. Although a slight haze may have been present, the overall
atmospheric conditions were highly favorable for EAS fluorescence
detection.

\begin{figure*}[t]
	\centering
	\fign{.84}{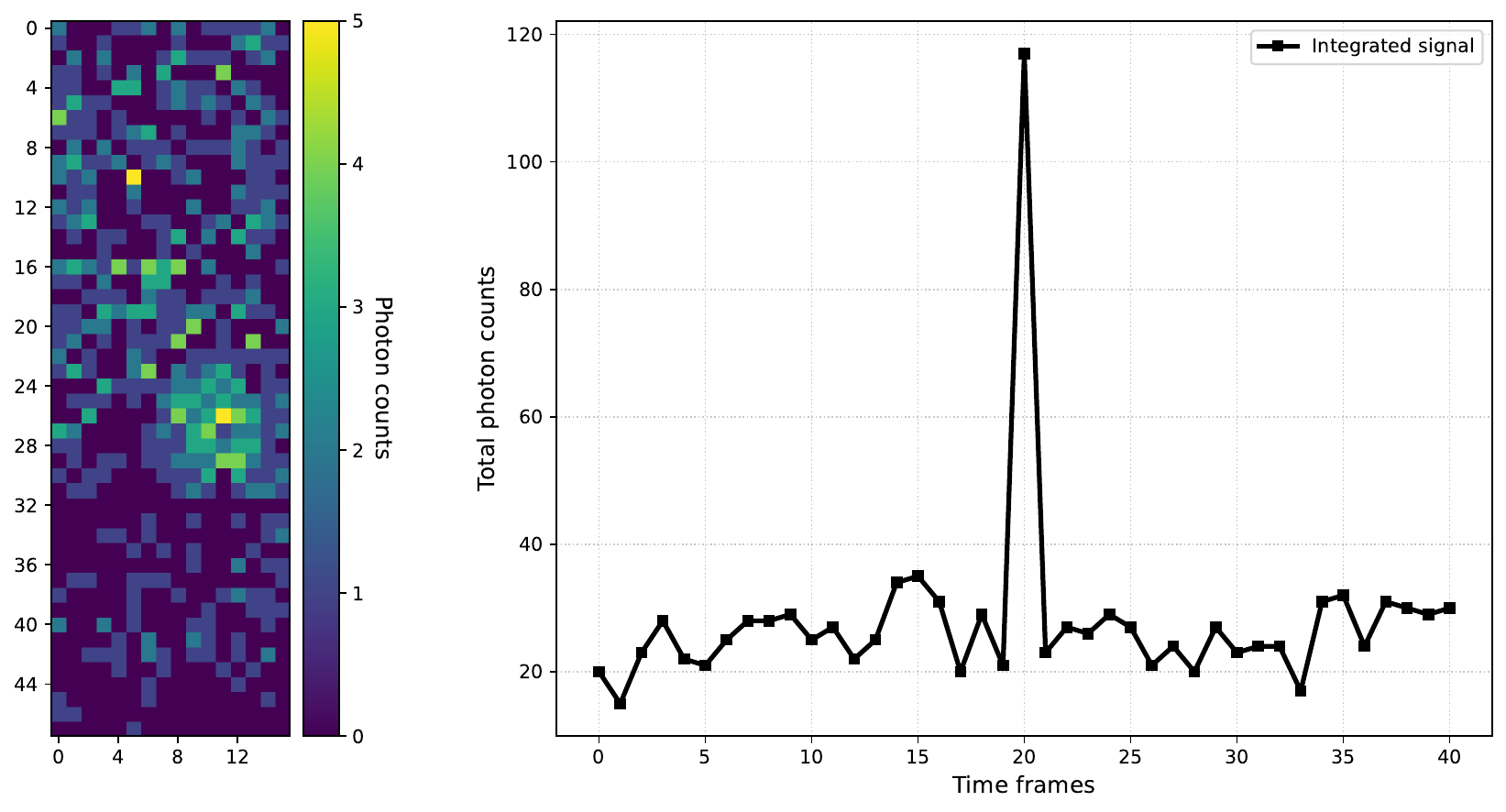}
	\fign{.82}{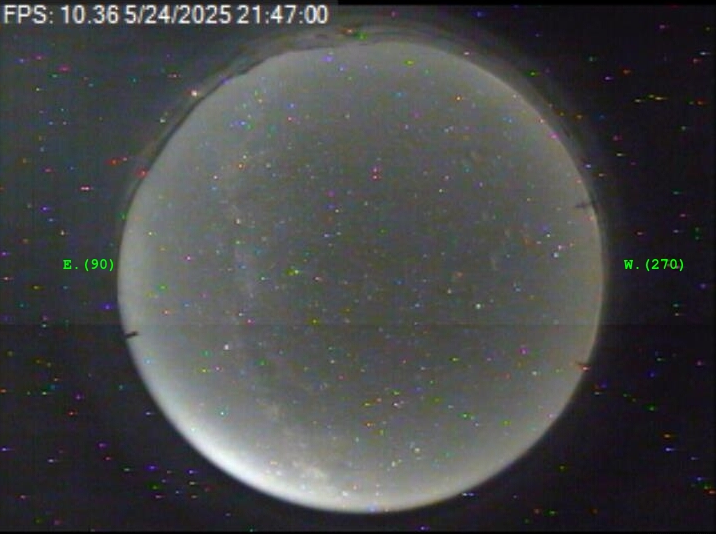}
	\caption{Top row: A track recorded on 24.05.2025 at 21:47:20 UTC.
	Similar to Fig.~\ref{fig:250423},
	the panels display the original SAFT signal (top left) and the
	integrated signal over the 40-bin readout window (top right).
	Bottom: all-sky camera image taken on 24.05.2025 at 21:47~UTC.
	}
	\label{fig:250524}
\end{figure*}

Figure~\ref{fig:250524} presents another example of a signal that we
attribute to an extensive air shower. It was recorded on May 24, 2025,
at 21:47:20 UTC, three nights prior to the new moon. Unlike the previous
track, the footprint of this signal on the photodetector does not
exhibit a pronounced elongated shape. As demonstrated by prior
simulations performed for the EUSO-TA telescope, such event topologies
can occur when the EAS axis is aligned closely with the telescope's
optical axis (line of sight).  The bottom panel of
Figure~\ref{fig:250524} displays an all-sky camera image captured 20~s
prior to the registration of this track. It is evident that the sky was
perfectly clear, to the extent that the Milky Way can be recognized.
Thus, the observational conditions at that moment can be considered
ideal for EAS fluorescence detection.

\begin{figure*}[!ht]
	\centering
	\fign{.7}{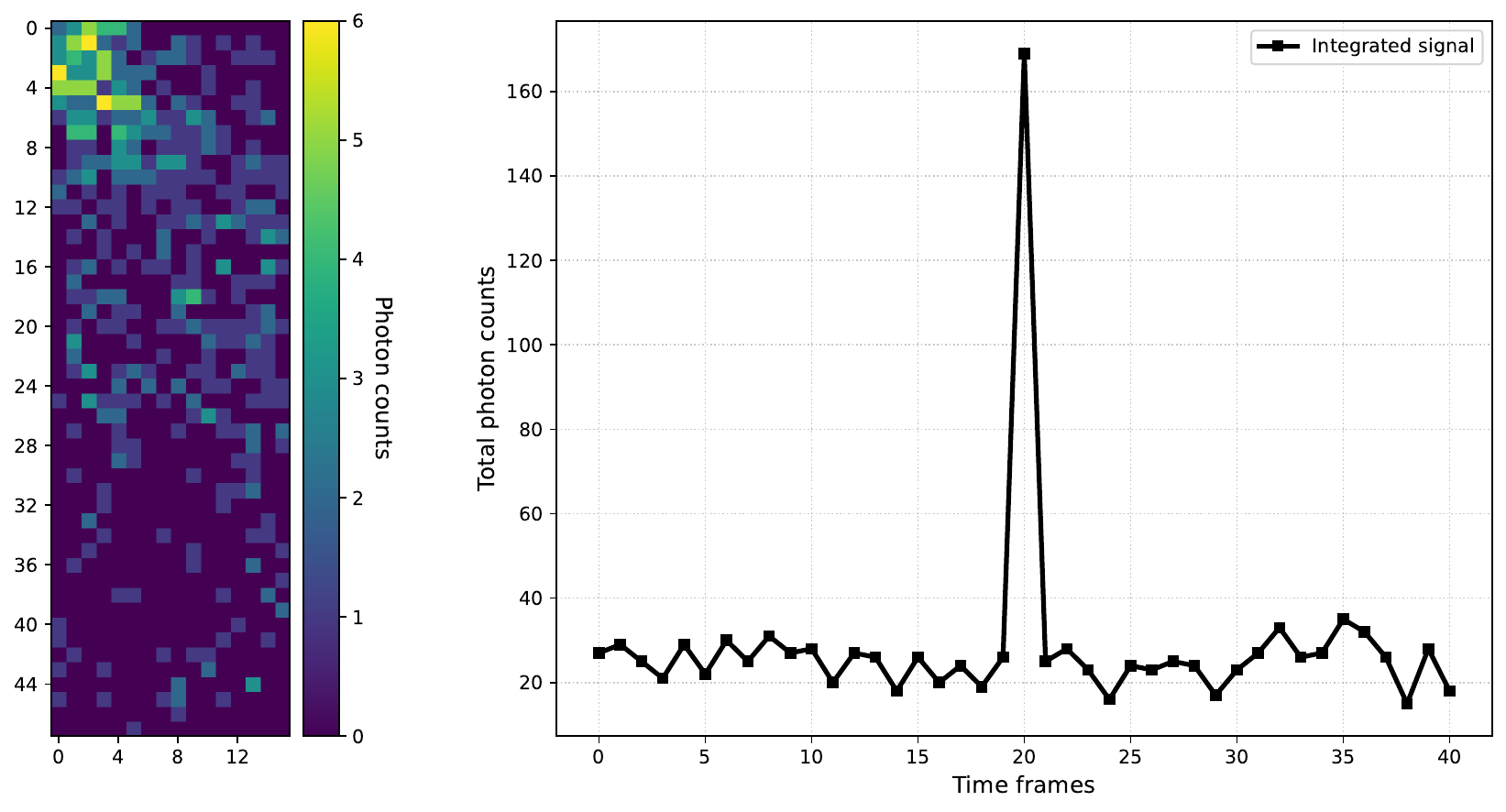}\\
	\fign{.7}{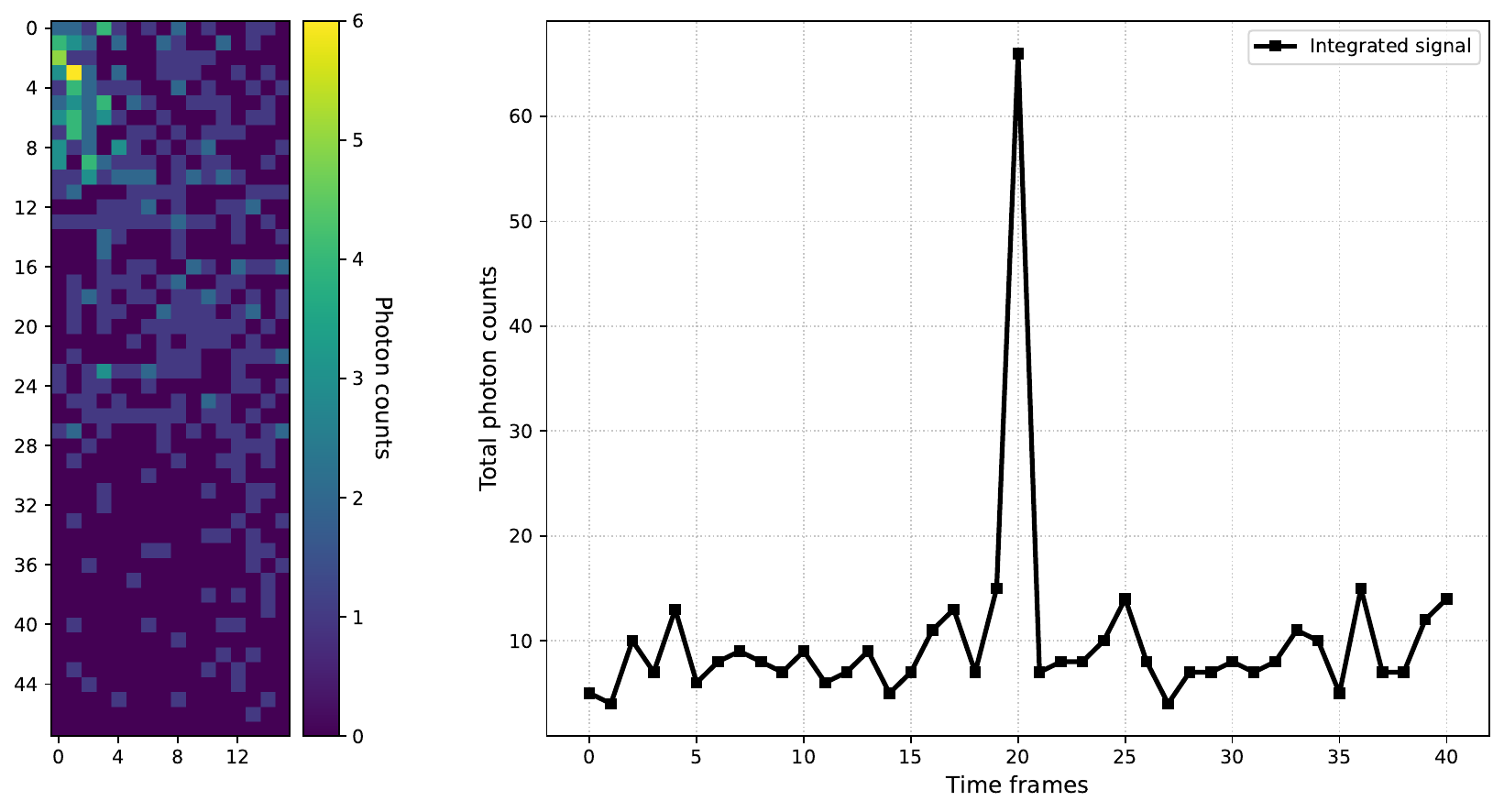}
	\fign{.7}{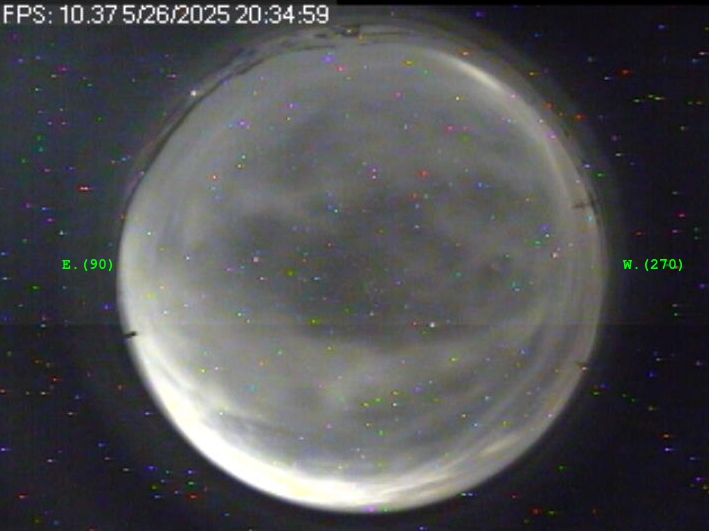}
	\caption{Top row: track image and temporal profile of the integrated
	pixel signal for the EAS recorded by the small fluorescence
	telescope on 26.05.2025 at 20:34:48 UTC.
	Middle row: corresponding plots for the simulated EAS event.
	Bottom: all-sky camera image taken on 26.05.2025 at 20:35~UTC.
	}
	\label{fig:250526}
\end{figure*}

\clearpage

Figure~\ref{fig:250526} shows yet another example of an EAS track,
recorded during the Mount Aragats observational campaign on May 26,
2025, at 20:34:48 UTC, on a moonless night. For comparison, the same
figure displays a simulated EAS track for the EUSO-TA telescope (shown
with a truncated focal surface), induced by a primary proton with an
energy of $\approx4$~EeV and a distance from the telescope to the shower
axis of approximately 10~km. The morphological similarity between the
focal plane images and the integrated temporal profiles of these two
showers—the real and the simulated one—is clearly visible.  {This
does not prove that the registered track originated from an extensive
air shower but provides evidence that this is possible.} The bottom
panel of Figure~\ref{fig:250526} shows an all-sky camera image taken a
few seconds after this track was recorded. While some cloud cover is
present (likely high-altitude cirrus clouds), the specific region of the
telescope's FoV where the signal was detected remains largely cloud-free
(cf. Fig.~\ref{fig:paips-fov}).

In total, within the completely analyzed subset of the observational
data, we have identified over 15 tracks that we attribute to extensive
air showers with a high degree of confidence. A dedicated software
simulation framework for the SAFT is currently under development, which
will enable a more rigorous and quantitative validation of these
findings.

\section{Discussion}

The analysis of data from the small-aperture fluorescence telescope
(25~cm lens diameter), acquired during observations at Mount Aragats on
clear nights near new-moon phases, has revealed several tracks induced
by extensive air showers crossing the instrument's field of view. To our
knowledge, this marks the first time EAS have been detected using a
fluorescence telescope with such a small aperture.

It is of fundamental interest to estimate the energies, or at least the
energy ranges, of the primary cosmic rays responsible for these EAS
events. Classical energy reconstruction techniques for fluorescence
telescopes are not designed to resolve this task using limited,
single-time-frame observations~\citep{auger-fd, ft-reco}. As demonstrated
by~\citet{Zotov-Trusov-2025}, obtaining such estimates may be feasible
using neural networks. However, training these networks necessitates a
highly accurate software simulation model of the telescope. At the time
of writing, such a model is not yet available, but its development
might be possible in the future.

The primary cosmic ray energy range can be roughly estimated through
indirect reasoning. Specifically, by knowing the number of detected EAS
events and determining the telescope's exposure, one could, in
principle, use the known CR energy spectrum~\citep{pdg} to estimate the
energy range of the recorded events. Unfortunately, this analytical
approach fails for FTs with a narrow field of view. As discussed in
detail in~\citet{eusota-2024} regarding EUSO-TA, the limited FoV implies
that the telescope frequently observes only a small segment of the EAS
longitudinal profile, which does not necessarily encompass the shower
maximum. Consequently, the ``visible'' energy appears lower than the
true primary energy, meaning that most EAS events whose true energies
exceed the instrumental threshold (for a given distance to the shower
axis) fail to trigger the detector. As a result, energy estimates
derived from the aforementioned indirect arguments are biased.
Nevertheless, extrapolating from the energies of CRs detected by
EUSO-TA~\citep{eusota-2024}, the results obtained by high-elevation FTs
in the Auger, Telescope Array, and LHAASO experiments~\citep{heat, tale,
lhaaso}, and factoring in the observation altitude, we can reasonably
expect the energy threshold for CRs detected by the SAFT to be in the
vicinity of $10^{17}$--$10^{18}$~eV.

Looking ahead, in addition to developing the SAFT simulation software and
performing energy reconstructions for the detected cosmic rays, we plan
to complete the analysis of the existing dataset and resume
observational campaigns in 2026. Furthermore, the construction of a
small reflecting telescope---a prototype for the ERA space mission---has
already commenced. The findings presented in this paper suggest that FTs
of this scale and design are viable not only for space-borne
applications but also for ground-based arrays. For instance, they could
provide independent calibration for the surface detectors of the
forthcoming TAIGA-100 experiment and perform measurements that
complement data from other detector systems.

\section*{Acknowledgments}

The authors express their sincere gratitude to their colleagues in the
JEM-EUSO collaboration, who kindly permitted the use of their
technologies developed earlier for the purposes of this study.
The work was carried out as part of a state assignment from Lomonosov
Moscow State University.
The authors thank the staff of the Cosmic Ray Division of the Yerevan
Physics Institute for assisting with the SAFT operation and data transfer
from Aragats to MSU.
The authors also acknowledge the assistance of an AI-based language
model in translating the manuscript and refining the English academic
style. The scientific content and conclusions remain the sole
responsibility of the authors.

\bibliographystyle{aasjournal}
\bibliography{first.bib}
\end{document}